\documentstyle[12pt,iopconf,epsfig]{article}
\begin{document}

\title{Gray models of convection in core collapse supernovae
\footnote{To appear in the proceedings of the Second Oak Ridge 
Symposium on Atomic and Nuclear Astrophysics}
}

\author{F. Douglas Swesty \footnote{E-mail: dswesty@ncsa.uiuc.edu}}
\affil{ Department of Astronomy and National Center for Supercomputing
Applications, University of Illinois at Urbana-Champaign, 
Urbana, IL 61801}

\beginabstract
\small
A major difficulty facing modelers of core collapse supernovae lies in
the need for an accurate description of the flow of neutrinos that
leak out of the post-collapse core.  In the interior of the collapsed
core the neutrinos have a short mean free path (MFP) and are diffusive
while in the exterior the MFP is much larger than the star and the
neutrinos free stream radially.  The greatest difficulty in modeling
the transport of neutrinos is to correctly describe the flow of
neutrinos in the intermediate regions where the neutrino distribution
function transitions between the two extremes.  In this region the
neutrino distribution function transitions from a Local Thermodynamic
Equilibrium (LTE) distribution to a non-LTE (NLTE) distribution.  The
difficulties of numerically modeling the neutrino transport in this
transition region have been compounded through the use of the gray
approximation to radiation transport.  The gray approximation assumes
that the neutrino distribution can be described by a distribution
function that is parameterized in terms of a neutrino temperature,
$T_\nu$, and a neutrino chemical potential, $\mu_\nu$.  However, these
parameters must be assumed. Furthermore, the parameters will also
differ between the LTE and NLTE regions.  Additionally, within the
gray approximation the location at which the neutrino distribution
function transitions from LTE to NLTE must be assumed.  By considering
a series of models where the LTE/NLTE decoupling point is varied we
show that the outcome of the numerical models is critically sensitive
to the choice of the decoupling point when the gray approximation is
employed.  We also examine the effects of the neutrino--electron
scattering (NES) rate which is difficult to correctly formulate within
the gray approximation.  We show that NES has a dramatic affect on the
the overall neutrino heating rate and the dynamics of the model.  This
result conflicts with the results of high resolution multi-group
models which model neutrino transport without employing the gray
approximation.

\normalsize

\endabstract

\section{Introduction}

For years researchers have realized that portions of the region of a
post-bounce core collapse supernova interior to a stalled prompt
shock are convectively unstable.  The origin of this convective
instability is fairly well understood.  After the ``bounce'' of the
collapsed stellar core the prompt shock wave is formed near the sonic
point which separates the inner homologous core from the supersonicly
infalling outer core.  As the shock begins to propagate outward from
this point (at an enclosed baryon mass of about $0.5-0.7 M_\odot$) 
it weakens as energy is expended in dissociating the heavy
nuclei into free nucleons.  This weakening of the shock produces a
negative entropy gradient that is convectively unstable.

Until recently, researchers had numerically modeled the prompt phase
of a core collapse supernova via 1-dimensional (1-D) radiation
hydrodynamic models.  While these models, in many cases, included
fairly sophisticated microphysics and radiation transport techniques
they were unable to accurately model the convection which is
inherently multi-dimensional in nature.  This situation changed
radically with the pioneering work of \cite{hbc92} who were the first
to conduct multi-dimensional simulations of this convection.  Within
the the past few years there have been a spate of models
\cite{mwm93,bhf95,jm96,mcbbgsu96a,mcbbgsu96b} that have begun to
carry out 2-D simulations of the evolution of this convectively
unstable region in the post-bounce epoch.  Most of these models have
shown convection to take place in the region between the
neutrinosphere and the stalled prompt shock.

However, the advancement to multi-dimensional numerical models of
post-bounce supernovae has forced some compromises in the way
that neutrino transport is modeled.  In this paper we report on some
consequences of the use of the gray approximation to radiation
transport in 2-D supernova models.  In the context of supernova models
the gray approximation has taken on a unique form.  The neutrinos in
the interior high density regions of the supernova are assumed to be
in LTE with matter while in outer regions of the core the neutrinos
are assumed to be thermally and chemically decoupled from the matter
\cite{cvb86} (hereafter CVB).  In the high density regions the
assumption of LTE is adequate.  However, the fact that the neutrinos
are decoupled in the exterior regions of the core require an
assumption of spectral properties for each species of neutrino.

The use of the gray approximation to describe the evolution of the
neutrinos in a core collapse supernova should be contrasted with the
multi-group treatment that has been employed in a plethora of 1-D
calculations \cite{mbhlsv87,mb89,bru85,bruenn89a,bruenn89b,slm94}.
The multi-group approach does not assume a spectral energy
distribution for the neutrinos, rather the spectrum is explicitly
modeled.  This is accomplished by solving a monochromatic transport
equation \cite{mm84} for a set of discrete energy ``groups'' which
span the spectral range of interest.

From 1-D simulations it has been known for some time that for the
problem of modeling neutrino transport in supernovae the gray methods
and the multi-group methods of neutrino transport can yield
substantially different results.  In particular we have discovered
that the results obtained by the use of the gray approximation are
extremely sensitive to the {\em ad hoc} choice of parameters needed to
describe the decoupling of neutrinos and matter and to the choices
made in the implementation of the weak interaction rates that describe
the coupling of neutrinos to matter.  Our purpose in this paper is to
describe these sensitivities and discuss how they effect the
multi-dimensional models in the convective epoch of supernovae.

\section{Numerical Models}


The numerical hydrodynamic algorithm employed in these calculations is
a modified version of the ZEUS-2D code (Stone \& Norman 1992) in
spherical polar coordinates.  The ZEUS-2D algorithm explicitly solves
the Euler equations of Newtonian hydrodynamics:
\begin{equation}
\frac{\partial \rho}{\partial t} + \nabla \cdot ( \rho {\bf v} ) = 0
\end{equation}

\begin{equation}
\frac{\partial ( e \rho)}{\partial t} + 
\nabla \cdot ( e \rho {\bf v} ) + (P+Q) \nabla \cdot {\bf v}
= \left(\frac{D e}{Dt} \right)_{source}
\end{equation}

\begin{equation}
\frac{\partial (v_i \rho) }{\partial t} + 
\nabla \cdot ( v_i \rho {\bf v} ) + \nabla (P+Q) + \rho \nabla \phi = 0
\end{equation}

This algorithm offers several advantages for this particular problem
(for details see \cite{swesty98f}).  We have made several major changes to
this algorithm to accommodate the unique nature of dense matter
hydrodynamics.  First, we have added a separate continuity equation to
treat the advection of electrons.  Second, we have modified the
ZEUS-2D algorithm to incorporate an arbitrary equation of state (EOS)
instead of the ideal gas assumed in \cite{sn92a}.  In doing so,
we have utilized the matter temperature $T$ as the fundamental
variable in solving the Lagrangean part of the gas energy equation
instead of using the internal energy.

A third major modification has been made in order to circumvent the
strong restriction on the timesteps imposed by the CFL stability limit
in polar coordinates near the origin.  We assume that the flow is
radial inside a certain radius, $r_{r}$ which eliminates the the
restriction posed by the angular part of the CFL criterion near the
origin.  For the calculations described in this paper we have employed
$r_{r} = 22.5$ km which is inside the proto-neutron star formed by the
core bounce.  In the calculations discussed in this paper we have
assumed that gravity is described by a spherically symmetric potential
which is easily calculated by integration over the domain of
computation.

Calculations were carried out in spherical polar coordinates with two
different mesh sizes.  The 1-D models were actually carried out using
the 2-D code with very low angular grid resolution, i.e. 3 angular
grid zones spanning the polar angular range of $(\pi/4,3\pi/4)$.  This
low angular grid resolution is sufficient to ensure that non-radial
flow features never develop.  The radial grid resolution spans three
separate ranges.  There are 60 radial zones covering the innermost 25
km of the grid.  Between 25 km and 400 km there are 132 zones, and
from 400 km to 900 km there are 25 zones that geometrically increase
in size with radius.  The 2-D models maintain the same radial zoning
but employ 90 angular zones over the same angular domain for an
angular zone width of $\Delta \theta = 1^\circ$


In the calculations described in this paper we have modeled the
evolution of three neutrino species: $\nu_e$, $\bar{\nu}_e$, and a
generic species $\nu_x$ which represents the $\nu_\mu$,
$\bar{\nu}_\mu$, $\nu_\tau$, and $\bar{\nu}_\tau$ neutrinos.
The evolution of the neutrino energy density in this model 
is described by the mixed-frame flux-limited diffusion equation:
\begin{equation}
\frac{D E_\nu}{D t} +
\nabla\cdot\left(D\nabla E_\nu\right) 
+E_\nu 
\left(
P_2 \frac{\partial v_r}{\partial r} +
Q_2 \frac{1}{\rho} \frac{D\rho}{Dt}
\right)
= - \left(\frac{D e}{Dt} \right)_{source}
\label{eq:rad_eng}
\end{equation}
where $E_\nu$ in this case refers to a generic neutrino energy density
that could be $E(\nu_e)$,$E(\bar{\nu}_e)$, or $E(\nu_x)$.  The $P_2$
and $Q_2$ coefficients involve the Eddington factor $\chi$ and are
described in \cite{mbhlsv87}.  The equation is implicitly finite
differenced on the same staggered mesh as the hydrodynamic code.  The
source term is linearized so that the entire implicit system of
transport equations is solved once per timestep.  Finally, in the
calculation described in this paper we have assumed that the neutrinos
diffuse only radially which decouples the solution of
(\ref{eq:rad_eng}) into a separate tridiagonal system for each angular
zone of the problem.  Our intent is not to mimic the variety of
schemes used in other works, but to employ a single scheme to
illustrate how the results of the simulations depend on the choice of
decoupling location.  We describe the details of the finite
differencing solution of equation (\ref{eq:rad_eng}) in
\cite{swesty98f}.

The diffusion coefficient $D$ is calculated by the Levermore--Pomraning
\cite{lp81} prescription. In order to calculate the source 
terms and diffusion coefficient of equation (\ref{eq:rad_eng}) the
gray approximation requires the assumption of a spectral distribution
for each neutrino species.  We discuss this choice in section 3.


For the equation of state we employ the finite temperature EOS of
\cite{ls91} using the Sk180 nuclear force parameter set
\cite{slm94}.  The EOS was tabularized using a thermodynamically
consistent interpolation scheme \cite{swesty96a} and is described in
\cite{swesty98f}.

The weak interaction rates were implemented in a fashion similar to
the two-fluid rates of CVB.  One exception to this are the electron
capture and neutrino capture rates.  These rates as implemented by
CVB, HBHFC, and BHF do not preserve detailed balance in the regions in
LTE.  We have not made the approximations employed in CVB in order to
simplify the phase space integrations but instead preserve detailed
balance by calculating the rates exactly by numerical quadrature.  The
other difference between our rates and the rates of the aforementioned
calculations is that we restrict the NES heating rates to reflect only
the down-scattering of neutrinos seen in multi-group calculations. If
this is not done the rates as implemented in the aforementioned
calculations can lead to an unphysical cooling of matter by NES in
high density regions.  Further details of the NES rate are discussed
in section
\ref{subsec:nes}.


For the initial conditions we collapsed a progenitor core using our
BRYTSTAR spherically symmetric multi-group radiation hydrodynamics
code \cite{slm94} which was run in Newtonian mode.  For the models
discussed here we employed the S15S7b core of \cite{wooweav95}.  This
particular progenitor model has a relatively small iron core.
The use of a multi-group code during collapse
and bounce allowed us to obtain post-bounce cores with accurate lepton
and entropy profiles.  We extract the initial data from the BRYTSTAR
runs at the point where the prompt shock begins to stall.  For the 2-D
models we apply a small $(\sim 1\%)$ sinusoidal perturbation to the
electron fraction in the convectively unstable region.

\section{Results}

\subsection{Choice of Decoupling Point}

In the interior of the collapsed stellar core the neutrino
distributions are in LTE because of the short MFP of the neutrinos.
The neutrinos in this region are in chemical and thermal equilibrium.
This permits a simple and accurate characterization of the neutrino
distribution as a Fermi--Dirac distribution where $T_\nu = T$ (the
matter temperature) and $\mu_\nu$ is given by the condition for beta
equilibrium.  The neutrino number density and energy density are
determined by $T_\nu$ and $\mu_\nu$.  In this situation the gray
approximation yields an excellent description of the evolution of the
neutrinos.

In contrast, in the exterior regions of the collapsed core the MFP
exceeds the radius of the star and the neutrinos do not interact with
matter.  Here the distribution function for the neutrinos reflects the
spectrum from where the neutrinos decouple from matter, i.e. the
spectrum ``freezes out'' in regions outside of the radius of the
decoupling point. The neutrino distribution at this point is not in
thermal or chemical equilibrium with the matter.  However, within the
gray approximation a distribution function must be assumed.  In the
case of the current generation of 2-D gray supernova models the
assumption has been that the distribution function in this NLTE region
can continue to be characterized by a Fermi-Dirac distribution
function.  However, the neutrino temperature and chemical potential in
this region {\it are not} given by the conditions for chemical and
thermal equilibrium with matter.  Rather, one must either assume that
$T_\nu$ and $\mu_\nu$ maintain the values that characterize the
distribution function at the decoupling point or make an {\it ad hoc}
assumption of their values.  In either case the neutrino number
density and energy density in the NLTE region are unrelated to $T_\nu$
and $\mu_\nu$.  These parameters serve only to characterize the
spectrum of the neutrinos and the matter-neutrino interaction rates.
With an {\it ad hoc} choice of parameters the sensitivities of the
models to the choice of parameters are clear: one can ``tune'' the
neutrino reheating rate in the gain region to yield an explosion 
(or not) for a given model.  For this reason we consider only
the case where the spectrum is assumed to be that of the decoupling 
point.

Unfortunately, the gray approximation does not constrain the location
of the point where the decoupling of neutrinos and matter occur.
Common lore is that decoupling occurs near the neutrinosphere, where
the neutrino optical depth is unity.  However, this notion is somewhat
ill defined for several reasons.  First, the neutrino opacity is a
strong function of neutrino energy, causing the physical location
of the neutrinosphere to vary with neutrino energy.  Thus the
location of the neutrinosphere is somewhat nebulous.  Secondly, it is
not clear what sources of neutrino opacity should be considered in
defining the neutrinosphere.  The iso-energetic neutrino--nucleon
scatterings which contribute strongly to the overall opacity
isotropize the neutrino distribution but do not thermalize it.  Thus
the surface of last-scattering is not the point of thermal
decoupling for the neutrinos.  The neutrino emission-absorption
reactions are not efficient at thermalizing the distribution because
of the large difference between the average neutrino energies and the
nucleon rest mass--energy.  Therefore, it is unlikely that the
neutrino distribution is thermalized exactly at the point where the
optical depth due to emission--absorption is unity.
Neutrino--electron scattering is a more efficient means of
thermalizing the neutrino distribution but the overall contribution of
NES to the opacity is small compared to the other interactions.  For
these reasons there is no clear physical criterion to determine where
decoupling should occur.

As a practical matter one can still employ criteria based on an energy
averaged optical depth or MFP to select a decoupling point in a gray
model \cite{hbhfc94,bhf95}.  Such criteria determine a ``gray
neutrinosphere'' based on the total opacity as calculated using the
grey spectrum.  But the question remains: To what extent does the
choice of the criterion affect the outcome of the model?  If the
choice of decoupling point, and thus the spectrum, in the NLTE regions
were robust the outcome of the numerical simulations should not
crucially depend on the particular choice.  As we have previously
mentioned there is no clear physical reason why any of the
aforementioned criteria should yield a precise location of the
decoupling point.  Accordingly, we examine in this paper how varying
these choices can affect the radiation--hydrodynamic models of the
convective region.
 
\subsection{Effects of Variations in the Decoupling Point}

In order to ascertain the sensitivity of the models to the choice of
decoupling point we carried out several 1-D and 2-D models where the
decoupling points were varied slightly in order to examine how the
outcome of the simulation would be affected.  The method we employ to
vary the location of the neutrino decoupling point is to choose the
decoupling point for each neutrino species based on the MFP, $\lambda$
for that species.  We select the decoupling point based on the
parameter $\xi=\Delta r/\lambda$.  By trial and error we have found
that varying $\xi$ between $0.5$ and $2$ we can cause the neutrino
decoupling point and spectrum to vary such that the location of the
gray neutrinosphere for the $\nu_e$ neutrinos varies over the range of
approximately $20-40$ Km.  The neutrinospheres for the $\bar{\nu_e}$
and $\nu_{\mu}$ neutrinos vary accordingly.  We briefly describe some
of our findings regarding the effects of this variation on the models.
For more detail the reader is referred to
\cite{swesty98f}.

For simplicity, we first consider three 1-D models where we choose the
decoupling point based on $\xi = 1/2$, 1, and 2.  Since $\xi$ is
defined in terms of the zone size there is no clear reason why one of
these choices should be preferred over another.  The neutrino
luminosities for these three models are displayed in Fig. 1.  Note
that the electron neutrino luminosities initially decrease by a factor of
approximately as $\xi$ is increased from $1/2$ to $2$.  
This difference gets larger later in the calculation as the $\xi=2$
model is overwhelmed by infalling matter driving the opacity up.
Note that the difference in the electron neutrino
luminosities between the $\xi=1/2$ and the $\xi=1$ cases is not nearly
as pronounced as it is between $\xi=1$ and $\xi=2$ cases.

The large change in neutrino luminosities between the various models
can be easily understood in terms of the rapid variation of the
temperature as a function of radius near the neutrinosphere.  A
typical example of this variation is shown in Fig. 2.  Since little
neutrino cooling has occured in the first few hundred milliseconds the
temperature profile is still largely determined by the hydrodynamic
evolution of the collapsed core during the prompt phase.  As $\xi$ is
decreased the decoupling occurs at a larger radius.  For the
timescales we are considering in this paper the decoupling radius for
the the electron neutrinos is approximately $23$ km for $\xi = 2$ and
approximately $39$ km for $\xi = 1/2$.  The corresponding electron
neutrino temperatures at the decoupling points are $T_{\nu_e} \approx
9.2$ MeV for $\xi=2$ and $T_{\nu_e} \approx 6.0$ MeV for $\xi=1/2$.
\begin{figure}[h]
\epsfig{file=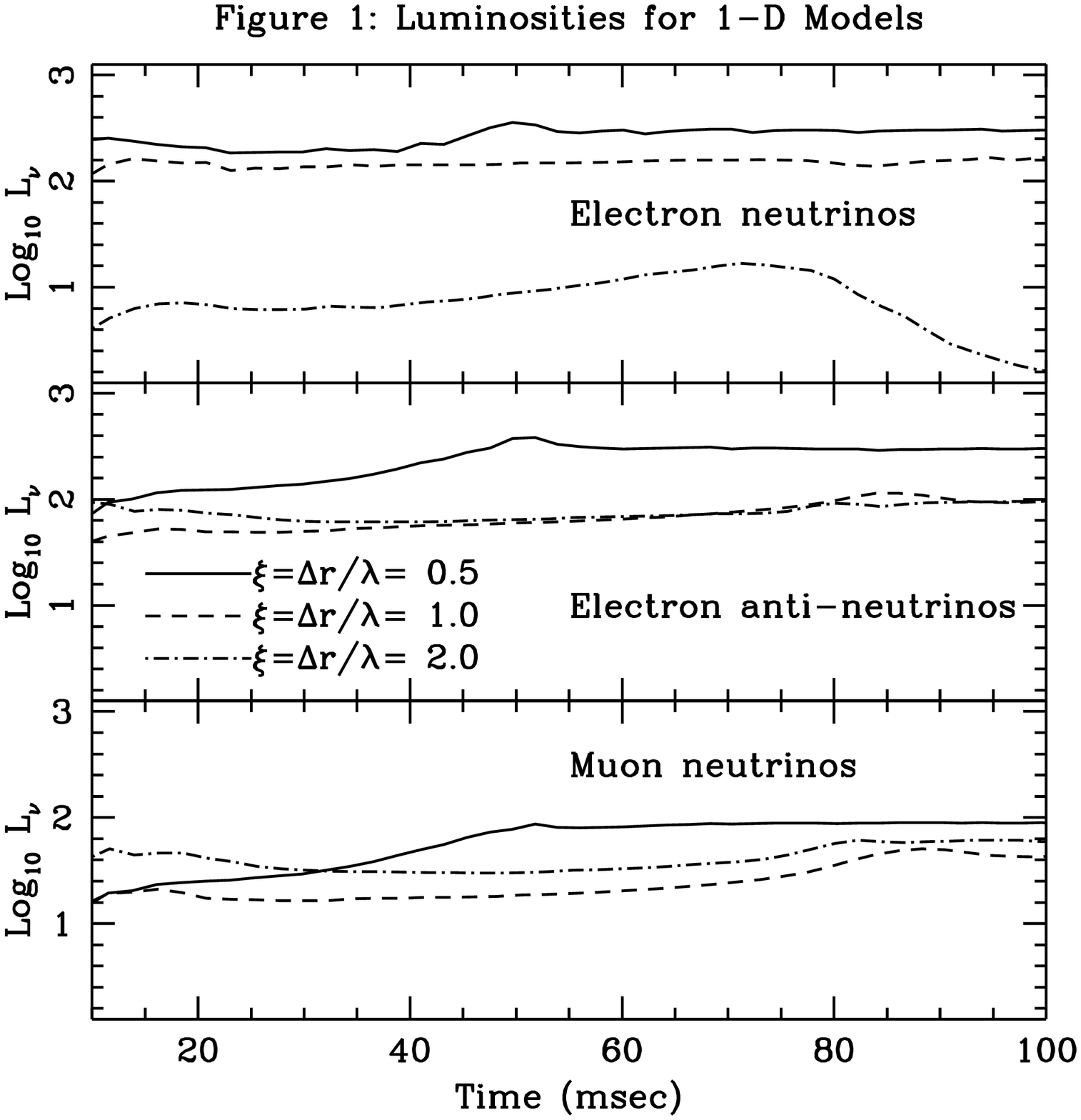,height=3.05in}
\epsfig{file=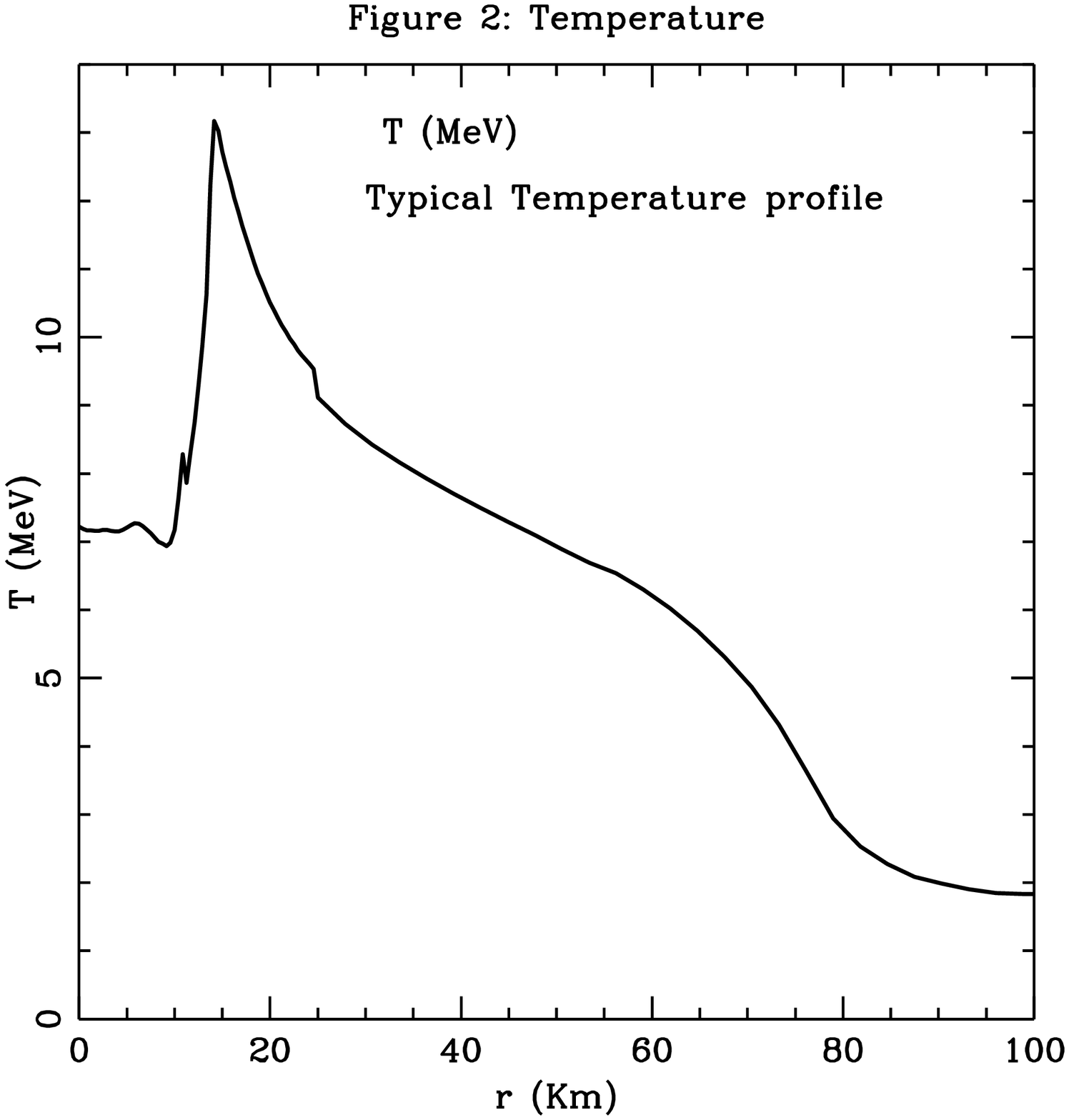,height=3.05in}
\end{figure}
The decoupling of the neutrinos at larger radii results in a decreased
electron neutrino temperature and a softer neutrino spectrum.  This
decrease in the electron neutrino temperature corresponds to a lower
average neutrino energy and in turn a lower opacity at a given
radius.  Consequently, a lower opacity results in a higher electron
neutrino luminosity.  A corresponding change is also reflected in the
electron anti-neutrino luminosities.  The radius of the neutrinosphere
for the electron anti-neutrinos decreases as $\xi$ is increased.  This
is most pronounced for the electron anti-neutrinos as the decoupling
cutoff is varied from $\xi=1/2$ to $\xi=1$ where the radius of the
neutrinosphere shifts from $39$ km to $22$ km.  The electron
anti-neutrino luminosity drops substantially with this shift.
However, as the cutoff is further shifted to $\xi=2$ there is little
change in the radius of the neutrinosphere for the electron
anti-neutrinos.  This is again reflected in the luminosities where the
$\xi=2$ case actually gives slightly higher luminosities then the
$\xi=1$ case for approximately $60$ milliseconds after core bounce.

In contrast to the electron neutrinos and electron anti-neutrinos the
$\mu$ and $\tau$ neutrinos (and anti-neutrinos) show smaller
variations in luminosity as the cutoff parameter is varied.  Again,
this can be attributed to the small variation in the $\nu_\mu$
neutrinosphere radii from $19.5$ km for $\xi=1/2$ to $17$ km for
$\xi=2$.  As expected the decoupling for the $\nu_\mu$'s occurs
deeper in the star.  The neutrino temperatures for the $\nu_\mu$
neutrinos in the NLTE region originate near the peak of the
temperature profile depicted in Fig. 2.  Small variations in the
decoupling point about this peak do not produce large variations in
the $\nu_\mu$ spectrum.
\begin{figure}[h]
\epsfig{file=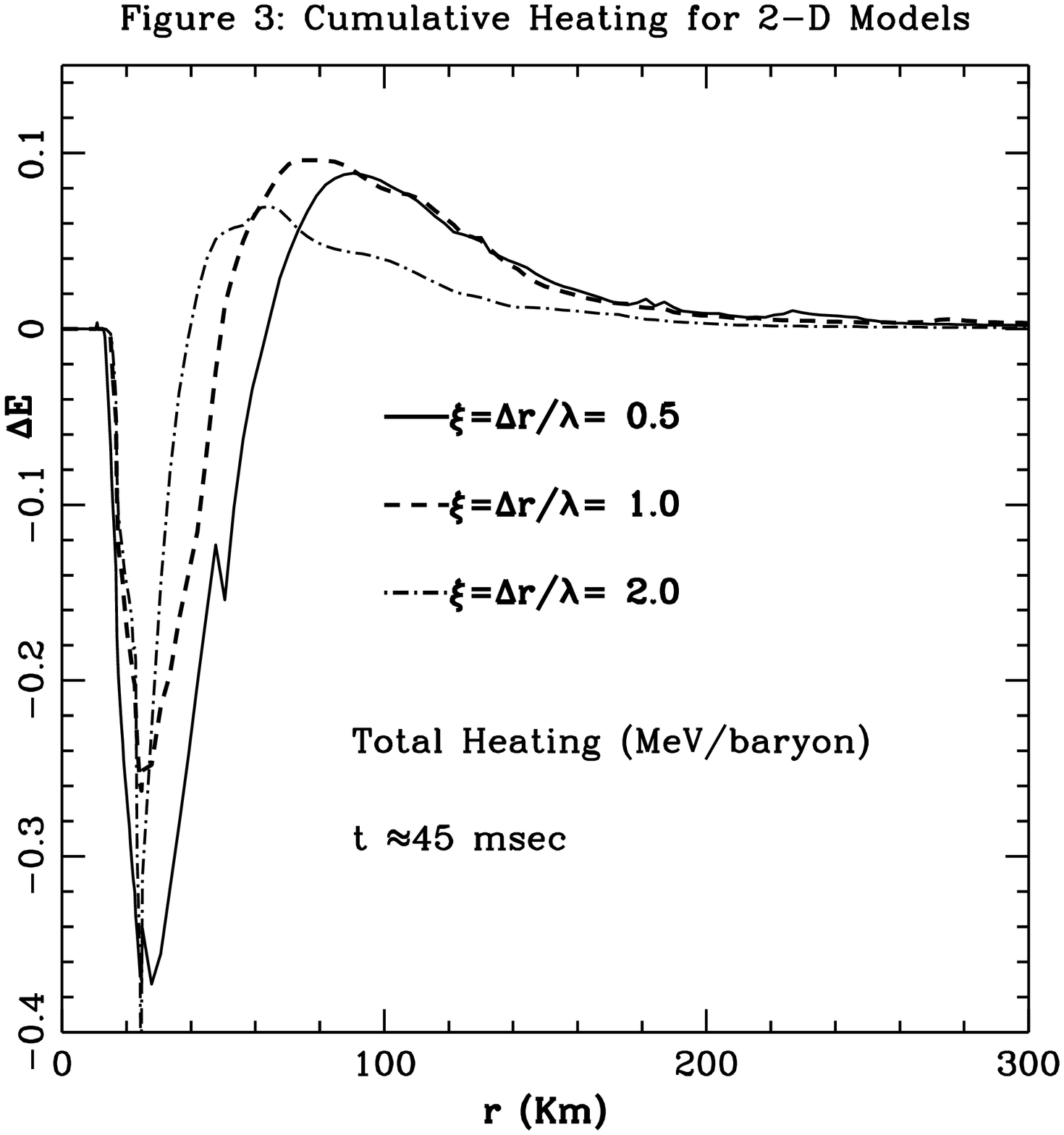,height=3.05in}
\epsfig{file=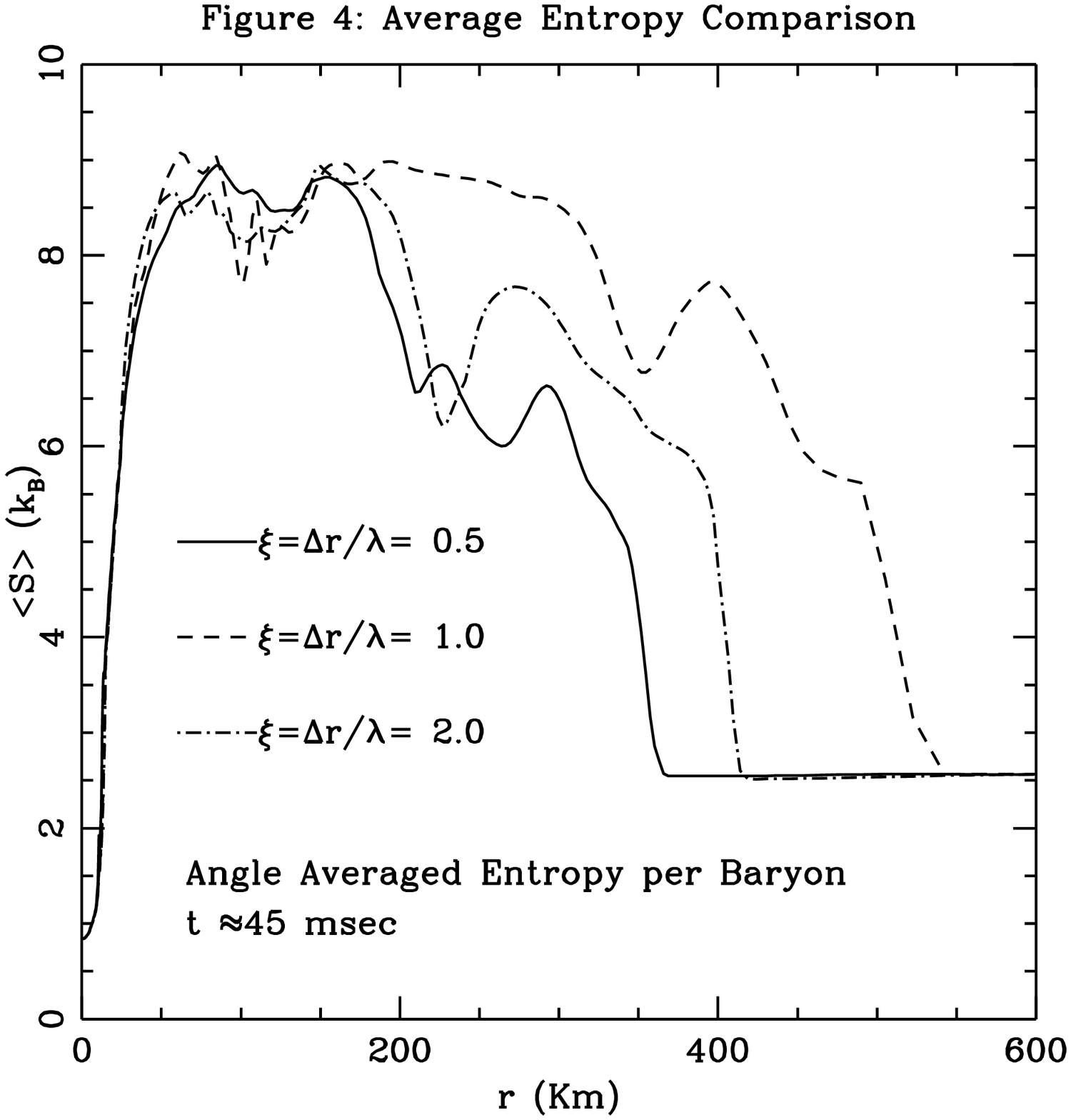,height=3.05in}
\end{figure}
The variation in the luminosities and average energies of the three
neutrino species also can have a substantial effect on the dynamics
and the chemical composition of the material in the region exterior to
the decoupling point.  This same effect is present in 2-D models where
the effects on the dynamics are more pronounced because of the
convection that occurs in these models.

The variations in the dynamical behavior of the 2-D models can be
understood in terms of the cumulative neutrino heating differences
among the three models.  The total amount of heating in the gain
region is shown in Fig. 3 at a time of approximately 45 milliseconds
after the beginning of the simulation when the convection is well
developed.  Note that the gain radius moves radially outward as $\xi$
decreases.  The luminosities increase as $\xi$ decreases, as in
the 1-D case, and the cumulative amount of heating in the gain region
increases.  Since the region from $r=60-200$Km is the base of the
convective region the increased heating in the gain region drives the
convection more vigorously.

While the peak heating rate in the outer regions for the $\xi=0.5$
case is comparable to that of the $\xi=1.0$ case the width of the gain
region is narrower than either the $\xi=1$ or $\xi=2$ cases, thus
resulting in the least total heating.  Consequently the $\xi=0.5$
model is the least vigorous of the three cases.  This can clearly be
seen in Fig. 4 where the entropy profiles are shown for the three
models.  Here the differences in the strength of the convection are
clearly revealed.  Given the cumulative heating profiles shown in
Fig. 3 it is not surprising that the $\xi=1$ case is showing the most
vigorous behavior.  By $t=90$ msec this model has exploded to the edge
of the grid.  In contrast, the narrow gain region of the $\xi=0.5$
case results in the the weakest convection of the three models.  By
$t=90$ msec the shock in the $\xi=0.5$ case actually recedes to a
radius of approximately $290$Km and shows signs of growing weaker.  It
is unclear whether this model will explode or not.  However, while the
$\xi=2$ model has a lower cumulative heating profile it has a broad
gain region and the model is slowly exploding.  Because of space
limitations we are unable to discuss on the effects of the decoupling
point choice on the chemical compositions.
 
We briefly mention that the variations in the spectra have a
substantial effect on the electron fraction in the convective region.
The composition of the material is set by the competing effects of
electron neutrino and anti-neutrino absorption in establishing kinetic
equilibrium in this material.  Thus the change in the neutrino spectra
and luminosities, as $\xi$ is varied, in turn yields large differences
in the electron fraction.  The reader is referred to
\cite{swesty98f} for a detailed discussion of this effect.

The wide variation among the three models clearly demonstrates the
sensitivity of the models to the choice of location of the LTE/NLTE
decoupling point. Such variation clearly points to the need to
eliminate the use of the gray approximation if we are to have
predictive models that do not employ free parameters.

\subsection{Role of the Gray NES Rate\label{subsec:nes}}

The formulation of the NES rate in the gray approximation is also a
source of substantial error in numerical RHD models of supernovae.  As
we previously mentioned NES is not a significant source of neutrino
opacity in the collapsed core but it has a strong effect on spectral
softening during core collapse.  However, through high resolution 1-D
multigroup calculations it is known that the NES heating rate in the
gain region during the post bounce epoch is dominated by the heating
rate due to absorption of neutrinos on nucleons (see Fig. 6 of
\cite{bruenn92}).  From this previous work we know that the 
inclusion of NES heating in post-bounce simulations should have little
effect on the dynamics of the model.  It is also known that the main
effect of NES is the spectral down-scattering of neutrinos thus
softening the spectrum and enhancing the overall luminosity.

The phenomenon of spectral softening is at odds with the {\it
a priori} choice of spectral parameters that must be made to employ
the gray approximation.  Once a spectrum has been assumed there is no
way to self--consistently account for the spectral rearrangement that
NES will cause.  Nevertheless, a number of attempts have been made to
estimate the heating rate due to NES within the gray approximation
\cite{cvb86,hbhfc94,bhf95}.  The reader is referred to these papers
for details of the formulation of these rates in the gray
approximation.  In some cases the formulations clearly have problems
as has been pointed out by CVB in \cite{cvb86}.

\begin{figure}[h]
\epsfig{file=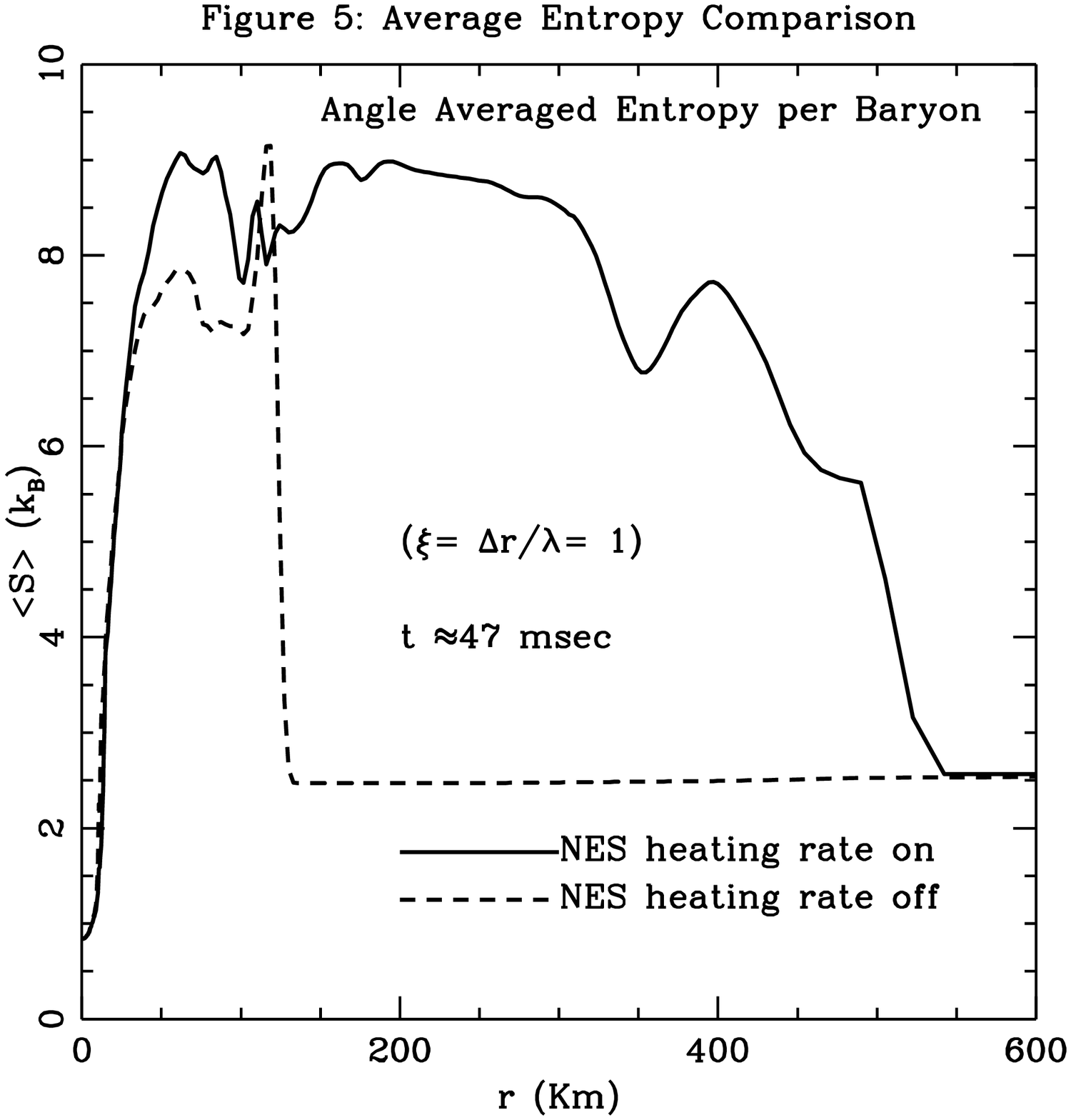,height=3.05in}
\epsfig{file=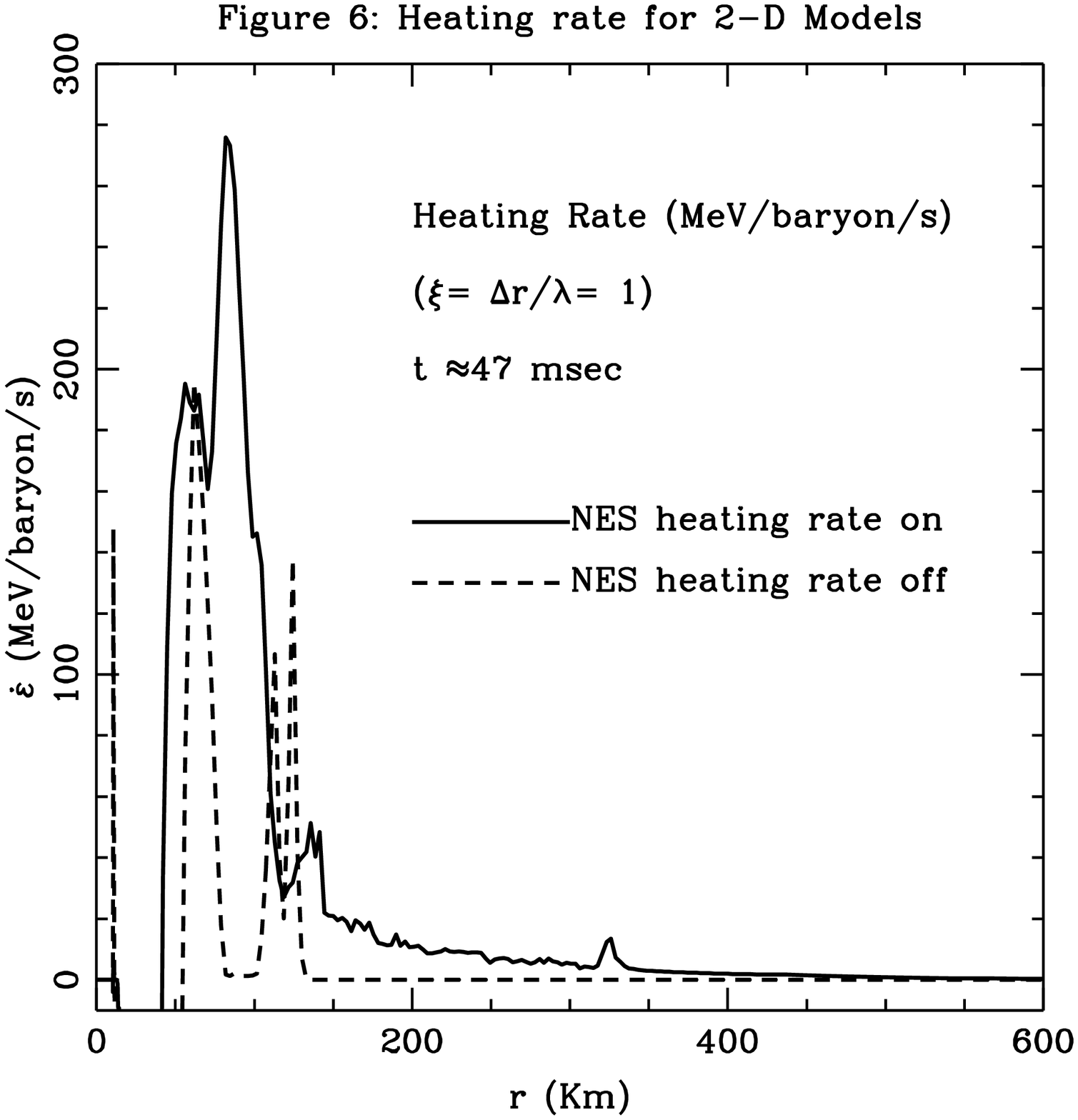,height=3.05in}
\end{figure}
In order to try to estimate the effect of these rates on the
convective models we carried out a comparison of the models with and
without the NES heating.  The results are dramatic.  The entropy
profiles for a $\xi=1$ model (the most robustly exploding 2-D model)
with and without NES are depicted in Fig. 5 at $t\approx 47$ msec.  In
the case without NES heating the explosion seems to ``fizzle'' for
this particular choice of decoupling point while it is extremely
vigorous for the case where NES heating is included using the CVB
rates.  The reason for this is clear from the heating rate profile
shown in Fig. 6.  The NES heating rate clearly gives a significant
contribution over a very large region exterior to the gain radius.
Clearly the excess heating in the region exterior to $r=150$Km caused
by the inclusion of NES heating is sufficient to completely alter the
outcome of the model.  At $t \approx 60$ msec when the model with NES
is well on the way to exploding, the shock in the model without NES
has receded slightly, as it becomes overwhelmed by infalling matter,
and seems to weaken.  This significant neutrino reheating of matter
due to NES conflicts with the results obtained by several multi-group
flux--limited diffusion
\cite{bru85,mbhlsv87,bruenn92} and multi-group Boltzmann
\cite{mezzb93c} calculations which have shown only minimal effects due
to NES.

The effects of the NES rate arise from the formulation of the rate in
terms of Fermi integrals over the neutrino and electron distributions
which does not correctly account for momentum exchange.  Furthermore,
gray formulations of this rate do not vanish in thermal equilibrium as
they should. CVB have made note of these difficulties and we refer the
reader to that work \cite{cvb86} for a thorough discussion of the
problem. We wish to point out that the bizarre behavior caused by NES
in the gray case indicates that it should not be included in these
simulations in its existing form.  These difficulties further indicate
the need to carry out future supernova models using multi-group
transport which allows for the neutrino--matter interaction rates to
be formulated more accurately.

\section*{Acknowledgments}
We would like to thank Steve Bruenn, Alan Calder, and Tony Mezzacappa
for many helpful discussions on these issues.  We wish to thank Stan
Woosley and Tom Weaver for the initial pre-collapse stellar models.  We
would also acknowledge financial support for this work under NASA ATP
grant NAG5-3099.  We also to like to thank NCSA and PSC for computing
support under NSF MetaCenter Allocation MCA975011.

\bibliographystyle{phjcp}
\small
\bibliography{astrojful,master}

\end{document}